\documentclass[aps,prb,twocolumn,superscriptaddress,bibnotes]{revtex4-1}
\usepackage{color}
\usepackage{graphicx}

\begin{document}


\title{Intertube effects on one-dimensional correlated state of metallic single-wall carbon nanotubes probed by $^{13}$C NMR}


\author{Noboru Serita}
\affiliation{Department of Physics, Graduate School of Science and Engineering, Tokyo Metropolitan University, Tokyo 192-0397, Japan,}
\author{Yusuke Nakai}
\email{nakai@tmu.ac.jp}
\affiliation{Department of Physics, Graduate School of Science and Engineering, Tokyo Metropolitan University, Tokyo 192-0397, Japan,}

\author{Kazuyuki Matsuda}
\affiliation{Institute of Physics, Faculty of Engineering, Kanagawa University, Yokohama 221-8686, Japan,}
\author{Kazuhiro Yanagi}
\affiliation{Department of Physics, Graduate School of Science and Engineering, Tokyo Metropolitan University, Tokyo 192-0397, Japan,}
\author{Yasumitsu Miyata}
\affiliation{Department of Physics, Graduate School of Science and Engineering, Tokyo Metropolitan University, Tokyo 192-0397, Japan,}
\affiliation{JST, PRESTO , 4-1-8 Hon-Chou, Kawaguchi, Saitama 332-0012, Japan,}
\author{Takeshi Saito}
\affiliation{Nanomaterials Research Institute, AIST, 1-1-1 Higashi, Tsukuba 305-8565, Japan}
\author{Yutaka Maniwa}
\email{maniwa@phyes.se.tmu.ac.jp}
\affiliation{Department of Physics, Graduate School of Science and Engineering, Tokyo Metropolitan University, Tokyo 192-0397, Japan,}

\date{\today}

\begin{abstract}
The electronic states in isolated single-wall carbon nanotubes (SWCNTs) have been considered as an ideal realization of a Tomonaga-Luttinger liquid (TLL). However, it remains unclear whether one-dimensional correlated states are realized under local environmental effects such as the formation of a bundle structure. Intertube effects originating from other adjacent SWCNTs within a bundle may drastically alter the one-dimensional correlated state. In order to test the validity of the TLL model in bundled SWCNTs, low-energy spin excitation is investigated by nuclear magnetic resonance (NMR). The NMR relaxation rate in bundled mixtures of metallic and semiconducting SWCNTs shows a power-law temperature dependence with a theoretically predicted exponent. This demonstrates that a TLL state with the same strength as that for effective Coulomb interactions is realized in a bundled sample, as in isolated SWCNTs. In bundled metallic SWCNTs, we found a power-law temperature dependence of the relaxation rate, but the magnitude of the relaxation rate is one order of magnitude smaller than that predicted by theory. Furthermore, we found an almost doubled magnitude of the Luttinger parameter.  These results indicate suppressed spin excitations with reduced Coulomb interactions in bundled metallic SWCNTs, which are attributable to intertube interactions originating from adjacent metallic SWCNTs within a bundle. Our findings give direct evidence that bundling reduces the effective Coulomb interactions via intertube interactions within bundled metallic SWCNTs.
\end{abstract}


\maketitle

\section*{I. Introduction}
A single-wall carbon nanotube (SWCNT) can be thought of as a graphene sheet rolled into a seamless hollow cylinder with a typical diameter in the nanometer range. A SWCNT can be either semiconducting or metallic, depending on the microscopic atomic arrangements and symmetry (chirality),~\cite{1} and as-prepared SWCNTs typically consist of mixtures of 2/3 semiconducting and 1/3 metallic tubes. The small diameter and high aspect ratio of a SWCNT lead to confinement of electrons along its circumferential direction, and thus a SWCNT is an ideal realization of a one-dimensional system.~\cite{2} Due to the one-dimensional nature of SWCNTs, electron-electron interactions are of great importance because the lack of screening enhances the effective Coulomb interactions between electrons. It has been theoretically shown that a long-range Coulomb interaction converts isolated metallic SWCNTs into a Tomonaga-Luttinger liquid (TLL),~\cite{3,4,5} in which the low-energy electron excitations are bosonic density waves of spin and charge.~\cite{6,7} In a TLL state, physical properties, such as the electronic density of states, and correlation functions exhibit power-law behavior, and their exponents quantify the strength of the electron-electron interactions. The strength of the effective electron-electron interactions is characterized by the charge Luttinger parameter (hereinafter referred to as Luttinger parameter), which ranges from 0 (very strong interactions) to 1 (no interactions, i.e., Fermi liquids) for repulsive interactions.~\cite{3,4,5} The theoretical estimate of the Luttinger parameter for an isolated SWCNT is approximately 0.2.

Although the formation of a TLL state in individual SWCNTs has been proposed, the effects of intertube interactions in a TLL state remain elusive, particularly to what extent the proposed TLL state is realized in an actual SWCNT sample. In fact, the electronic properties of SWCNTs are extremely sensitive to their environmental surroundings.~\cite{8,9,10} It is well known that SWCNTs pack together closely and form a triangular lattice called a bundle structure,~\cite{11,12} which can also give rise to environmental effects.~\cite{13,14,15,16,17} 
For example, bundling of SWCNTs produces a redshift of the optical transition energy, and has been understood as the consequence of the mutual dielectric screening provided by adjacent SWCNTs in a bundle, which reduces the electron-electron repulsion in each SWCNT.~\cite{18,19} Thus, effective Coulomb interactions between adjacent metallic SWCNTs within a bundle of metallic SWCNTs may be largely screened out, leading to a significant deviation from the proposed TLL state. 
In the case of bundled mixtures of semiconducting and metallic SWCNTs, the first photoemission spectroscopy (PES) experiment,~\cite{Ishii} which evidenced the TLL state, reported the same Luttinger parameter as the theoretical estimate for an isolated SWCNT, which was subsequently confirmed by other group.~\cite{22}
Contrary to the above expectation of the strong dielectric screening due to metallic SWCNTs within a bundle, previous PES experiments on bundled metallic SWCNTs~\cite{20} identified a TLL state which has a comparable Luttinger parameter (= 0.21) to that of bundled mixtures of semiconducting and metallic SWCNTs (= 0.19).~\cite{Ishii,22} Ayala {\it et al}. concluded that the interaction between different metallic SWCNTs within a bundle of only metallic SWCNTs is small enough to stabilize a TLL state.~\cite{20} The results of transport experiments in isolated and bundled SWCNTs can also be interpreted in terms of a TLL state,~\cite{23,24,25} although there remain difficulties in interpreting the role of the contacts. Thus, the effect of bundling on the one-dimensional electronic state in SWCNTs remains elusive. Understanding the changes in the physical properties of SWCNTs due to bundling is very important not only from the perspective of fundamental physics but also for the device application of SWCNTs, because SWCNTs naturally form bundles in typical synthesis processes. An understanding of intertube interactions due to bundling would allow the full potential of SWCNTs to be exploited for future nano-electronics.

Nuclear magnetic resonance (NMR) is a useful probe for characterizing low-energy electronic states, such as the density of states near the Fermi energy, without electrical contacts, and may shed new light on the one-dimensional electronic states of SWCNTs. An advantage of NMR is that it is a bulk-sensitive probe in contrast to surface-sensitive PES experiments. Previous NMR measurements of bundled mixtures of metallic and semiconducting SWCNTs showed a clear power-law temperature dependence of the nuclear spin-lattice relaxation rate $T_1^{-1}$,~\cite{26} which is consistent with a TLL state. 
Previous NMR study on inner tubes of bundled double wall carbon nanotubes~\cite{Singer} was also analyzed in the framework of a TLL state.~\cite{Dora}
However, it has not yet been resolved why the obtained Luttinger parameter of the bundled mixtured SWCNTs (= 0.34)~\cite{26} is approximately 70\% larger than the theoretical estimate and values from transport and PES experiments. 

The purpose of this paper is to investigate the electronic excitation microscopically by NMR to test the validity of the TLL model in bundled SWCNTs. For this purpose, we used two different samples, one a mixture of metallic and semiconducting SWCNTs, and the other consisting of highly concentrated metallic SWCNTs, where the adjacent environment of the metallic SWCNTs may differ significantly by the formation of a bundled structure.

\section*{II. Experiment}
We used two different pristine SWCNTs produced by e-DIPS (enhanced direct injection pyrolytic synthesis)~\cite{27} and an arc-discharge method. In the e-DIPS sample, $^{13}$C isotopes were enriched to more than 90\% in order to increase the NMR signal intensity, and no metallicity sorting was performed. The highly concentrated metallic SWCNTs (more than 95\%) were prepared from Arc-SO SWCNT soot (Meijo Nano-carbon) by a density gradient ultracentrifugation technique.~\cite{28} The concentration of the metallic fraction was determined by optical absorption (Shimadzu UV-3600).~\cite{29} In order to reduce ferromagnetic impurities, we performed an acid treatment and applied a density gradient ultracentrifuge method, as described in our previous paper.~\cite{30} After the treatment, the samples were washed with ethanol and acetone. Vacuum filtration was then performed to prepare buckypaper with a thickness of approximately 100 $\mu$m. The sample was finally annealed under a dynamic vacuum at 500 $^{\circ}$C in order to avoid doping from adsorbed gas and molecules. The obtained NMR samples were sealed in a quartz tube filled with 100 Torr of high purity helium gas. The average diameters of the SWCNT samples were characterized by powder X-ray diffraction (XRD) at the BL8B station in the Photon Factory Facility, KEK, Japan. The DC magnetic susceptibility was measured using a SQUID magnetometer (MPMS, Quantum Design). The SQUID measurements indicated that the amount of ferromagnetic impurities in the purified sample was two orders of magnitude smaller than in pristine material. $^{13}$C NMR measurements were performed under a magnetic field of 9.4 Tesla. The nuclear spin-lattice relaxation rate $T_1^{-1}$ was determined by a saturation recovery method. We found that the recovery curve for the nuclear magnetization did not show a single exponential form, but could be fit to a stretched exponential form, as reported in a previous paper,~\cite{26} with a stretching exponent $\beta$ = 0.82 for the bundled mixed sample and $\beta$ = 0.60 for the bundled metallic sample, both of which were temperature independent within experimental accuracy.

\section*{III. Results and discussion}
Figure 1 shows a $^{13}$C NMR spectrum of the $^{13}$C-enriched e-DIPS sample, which consists of bundled mixtures of metallic and semiconducting SWCNTs. In contrast to the usual $^{13}$C NMR spectrum observed for our unenriched metallic SWCNT sample,~\cite{31} the spectrum in Fig.~1 forms a doublet. Because the e-DIPS sample is almost entirely $^{13}$C enriched, we attributed the splitting to the nearest neighbor $^{13}$C nuclear dipolar interaction. We found from NMR measurements under a magnetic field of 4 Tesla that the splitting is magnetic field independent (not shown), which is one of the characteristics of the well-known Pake doublets.~\cite{32} From the splitting we can estimate the carbon-carbon (C-C) bond length as follows. Here, we neglect the curvature of a graphene sheet and assume that the local structure around a $^{13}$C atom with three $^{13}$C nearest neighbors is a hexagonal-honeycomb plane for simplicity. The central $^{13}$C nuclei at the origin (0, 0) experiences a dipolar field from the three nearest neighbor $^{13}$C nuclear spins at ($r$, 0), ($-r/2$, $r\sqrt{3}/2$), and ($-r/2$, $-r\sqrt{3}/2$) on a graphene hexagonal sheet, where $r$ is the C-C bond length. These $^{13}$C atoms can thus be thought of as three Pake pairs. Neighboring Pake pairs on a hexagonal honeycomb lattice have been previously reported for ThAl$_2$ powder samples, and the Al interatomic distance was determined from NMR spectral doublets.~\cite{33} The splitting corresponds to $3\nu_{\rm D}$, where $\nu_{\rm D} = \frac{3Ih}{2r^3}(\frac{\gamma}{2\pi})^2$, $I = 1/2$, $\gamma =10.705\times 2\pi$ MHz/T. The obtained C-C bond length is 1.34 \AA, which shows reasonable agreement with our recent XRD result.~\cite{34} Thus, $^{13}$C enriched NMR can be applied for the determination of the C-C bond length of SWCNTs.
\begin{figure}
\centering
\includegraphics[width = 7.5cm]{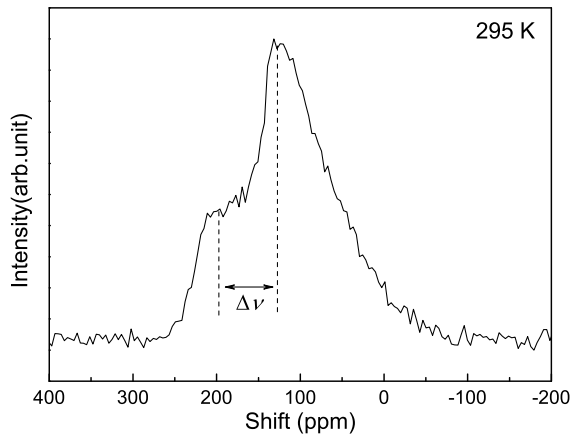}
\caption{$^{13}$C NMR for bundles of mixed semiconducting and metallic SWCNTs at 295 K under 9.4 Tesla.\label{fig1}}
\end{figure}

A noticeable characteristic of the TLL is a power-law dependence of the physical properties with a critical exponent describing the strength of the effective Coulomb interactions. The NMR relaxation rate $T_1^{-1}$ is known to be a useful probe for the electronic spin correlation functions. Figure 2 shows the temperature dependence of $(T_1T)^{-1}$ of bundled SWCNT samples with different metallicity (closed squares and diamonds), as well as the previous results for bundled mixtures of metallic and semiconducting SWCNTs (open circles).~\cite{26} All the SWCNT samples exhibit a power-law temperature dependence of $(T_1T)^{-1}$ above 20 K. Note that $(T_1T)^{-1}$ crosses over to a gap-like behavior at low temperatures, as reported previously.~\cite{26} The origin of this behavior is beyond the scope of the present study. The power-law temperature dependence of $(T_1T)^{-1}$ indicates the realization of a TLL state. It is noteworthy that the absolute values of $(T_1T)^{-1}$ and the temperature exponents differ among the samples. Following Ref.~\cite{26}, we estimated the Luttinger parameter as 0.18 $\pm$ 0.03 for the bundled mixtures of metallic and semiconducting SWCNTs (closed squares). The value of the obtained Luttinger parameter is almost the same as that derived theoretically,~\cite{3,4,5} and those obtained from transport~\cite{23,24} and PES measurements for bundled mixtures of metallic and semiconducting SWCNTs.~\cite{Ishii,22} In contrast, $(T_1T)^{-1}$ for the bundled metallic SWNCTs (closed diamonds) is one order of magnitude smaller than that for the bundled mixtures of metallic and semiconducting SWCNTs. This indicates that bundling of metallic SWCNTs suppresses the low-energy spin excitation significantly. Furthermore, we estimated the Luttinger parameter for the bundled metallic SWCNTs as 0.38 $\pm$ 0.03, which is approximately two times as large as that for the bundled mixtures of metallic and semiconducting SWCNTs. Because the Luttinger parameter is a measure of the electron-electron repulsion, this larger parameter indicates that effective Coulomb interactions are weaker in bundled metallic SWCNTs than in the bundled mixtured sample.  To the best of our knowledge, this is the first evidence that bundling of metallic SWCNTs weakens the electron-electron interaction strength while bundling of mixtures of metallic and semiconducting SWCNTs does not alter it. 
\begin{figure}
\centering
\includegraphics[width = 7.5cm]{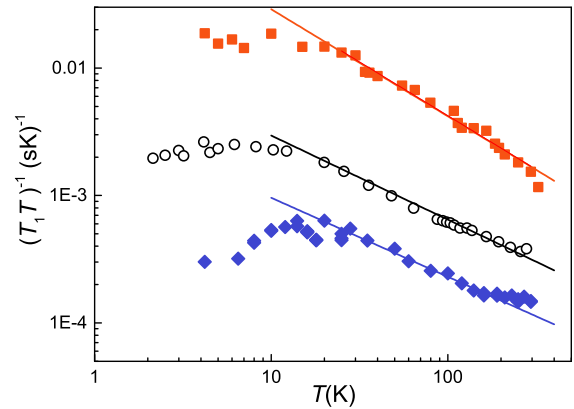}
\caption{Temperature dependence of $(T_1T)^{-1}$ for bundled SWCNT samples with different metallicity measured under 9.4 Tesla. The closed squares represent bundles of mixed metallic and semiconducting SWCNTs. The closed diamonds represent bundles of highly concentrated metallic SWCNTs. The open circles represent bundles of mixed metallic and semiconducting SWCNTs from Ref.~\cite{26}. The solid lines represent fits to $T^{\alpha}$.\label{fig2}}
\end{figure}

Recently, Kiss {\it et al}. demonstrated that the magnitude and temperature dependence of $(T_1T)^{-1}$ reported by Ihara {\it et al}. can be explained quantitatively in the framework of TLL theory.~\cite{26,35} They found that the NMR relaxation rate is enhanced by orders of magnitude compared to a Fermi liquid with the same density of states (DOS) due to a TLL state.~\cite{35} They derived the equation
\begin{equation}
 (T_1T)^{-1} = A_{\rm eff}^2\frac{k_{\rm B} C(K)}{\hbar} \left( \frac{2\alpha \pi^2}{\Xi(n,m)} \right)^K [N(E_{\rm F}) k_{\rm B} T]^{K-2}N(E_{\rm F})^2,
\end{equation}
where $A_{\rm eff}$ is the effective hyperfine coupling constant for $^{13}$C, (n, m) indicates the SWCNT chirality, $N(E_{\rm F})$ is the density of states at the Fermi energy, $K$ is the temperature exponent of $(T_1T)^{-1}$ determined by experiment, $\alpha$ is a constant depending slightly on the chirality (see in Ref.~\cite{35}), $\Xi(n,m)=\sqrt{3}/2\sqrt{n^2+nm+m^2}$ and $C(K) = \sin({\pi K})\Gamma(1-K)\Gamma^2(K/2)/2$. The dotted lines in Fig.~3 represent the calculated $(T_1T)^{-1}$ using the parameters listed in the table. Note that in order to compare the calculated values with the experimental result, we consider spin-diffusion effects. The observed $T_1^{-1}$ for bundled mixtures of metallic and semiconducting SWCNTs may be considered to be an average value due to spin diffusion among the 1/3 metallic and 2/3 semiconducting SWCNTs. Therefore, the $T_1^{-1}$ plotted in Fig.~3~(a) is taken to be three times larger than the observed $T_1^{-1}$ for bundled mixtures of metallic and semiconducting SWCNTs plotted in Fig. 1, because undoped semiconducting SWCNTs should exhibit a much longer $T_1$ than metallic SWCNTs. Note that our previous thermoelectric measurements suggest that the Fermi level for as-prepared SWCNT films lies within the band gap.~\cite{37,38} As shown in Fig.~3~(a), the relaxation rate calculated in the framework of the TLL theory reproduces the experimental results for bundled mixtures of metallic and semiconducting SWCNTs quite well. In contrast, the experimental $(T_1T)^{-1}$ for the bundled metallic SWCNT sample is one order of magnitude smaller than the theoretically calculated value (see Fig.~3~(b)). The correlated electronic states for the bundled metallic SWCNTs are qualitatively consistent with the TLL theory, as is obvious from the power-law behavior of $(T_1T)^{-1}$, but the low-energy electronic spin excitations are quantitatively different from those predicted by the TLL theory (dashed line in Fig.~3~(b)) due to the reduced Coulomb interactions. 
\begin{table*}[htb]
\begin{center}
\begin{tabular}{|c|c|c|c|c|c|} \hline
                                          & $A_{\rm eff}$ (eV) & $K$  & (n, m)   & $N(E_{\rm F})$ (states/eV/spin/atom) & $\alpha$ \\\hline\hline
Mixed ($d=2.1$ nm)                     & $3.6\times10^{-7}$ & 1.18 & (15, 15) & 0.00485                              & $2.39\times10^{-3}$ \\\hline
Metal ($d=1.4$ nm)                        & $3.6\times10^{-7}$ & 1.38 & (10, 10) & 0.0073                               & $3.58\times10^{-3}$ \\\hline
Mixed ($d=1.6$ nm, Ihara {\it et al}.) & $3.6\times10^{-7}$ & 1.34 & (12, 12) & 0.006                                & $2.98\times10^{-3}$ \\\hline

\end{tabular}
\end{center}
\caption{Parameters used for the calculation of Eq.~(1). $N(E_{\rm F})$ data were taken from Ref.~\cite{36}.}
\label{}
\end{table*}
\begin{figure*}[htb]
\centering
\includegraphics[width = 15cm]{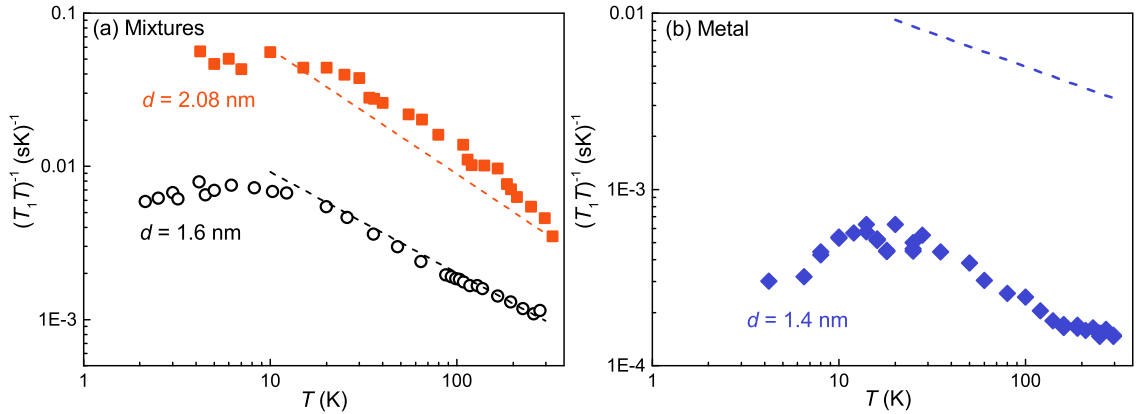}
\caption{Comparison between the experimental $(T_1T)^{-1}$ and that calculated by TLL theory for (a) bundled mixtures of metallic and semiconducting SWCNTs and (b) bundled metallic SWCNTs. Note that the $(T_1T)^{-1}$ value in (a) is three times larger than those in Fig.~1 because of spin-diffusion averaging (see text). The calculated results based on Ref.~\cite{35} reproduce the experimental $(T_1T)^{-1}$ quite well in (a).\label{fig3}}
\end{figure*}

We ascribe the reduced Coulomb interaction observed in the bundled metallic SWCNTs to intertube interactions between neighboring SWCNTs. Two types of intertube interactions have been discussed,~\cite{16,17,18,19} one of which is the direct coupling of the electronic states in adjacent SWCNTs (orbital hybridization effects), which may impart a three-dimensional nature to the electronic states of SWCNTs. The second type of interaction is dielectric screening induced by other adjacent nanotubes, leading to weakening of the Coulomb interactions. Dielectric screening has been proposed as a more robust and likely mechanism of intertube coupling than direct coupling.~\cite{18,19} Considering the larger dielectric constant for a metallic SWCNT than for a semiconducting SWCNT,~\cite{39,40} we conjecture that dielectric screening is a more natural explanation for our NMR results, although the possibility of direct coupling cannot be fully disregarded.

We now discuss the different Luttinger parameters for bundled metallic SWCNTs observed in our NMR and previous PES measurements.~\cite{20} The Luttinger parameter obtained by PES experiments (= 0.21) is comparable to that for the bundled mixtures of metallic and semiconducting SWCNTs,~\cite{20,Ishii,22} but is almost half that of our NMR estimate (= 0.38). This difference can be ascribed to the fact that PES measurements are surface sensitive; the mean free path of a photoelectron in the experimental energy range is several \AA, which is two orders of magnitude smaller than a typical bundle size (order of 10 nm).~\cite{41} Therefore, PES experiments can only probe the bundle surface where intertube interactions are weakened due to the smaller number of adjacent SWCNTs than in the bulk of the bundles probed by $^{13}$C NMR, for which we estimate a skin depth of 50 $\mu$m considering the electronic conductivity of a single bundle.~\cite{42}

Although the origin of the weakened Luttinger parameter for bundled mixtures of metallic and semiconducting SWCNTs observed in Ref.~\cite{26} remains unknown, it might be attributed to an enhanced defect concentration in the sample, which was produced by laser ablation with nonmagnetic Rh and Pt as catalysts.~\cite{43} This method has a disadvantage of giving a low yield of SWCNTs, which leads to a large defect concentration. The GD ratio for the sample inferred from Raman scattering experiments was reported to be as high as 20, whereas it was more than 100 for the e-DIPS sample.~\cite{27} For such defective SWCNTs, the average tube length effectively decreases, which leads to a larger Luttinger parameter according to Eq.~(3.23) in Ref.~\cite{5}. Further studies are needed to clarify the effects of defect on the electronic correlation in a TLL state for SWCNTs, which may open up a new method for manipulating nanotube properties. 

\section*{IV. Conclusion}
In conclusion, the NMR measurements clarified contrasting one-dimensional electronic states that depend on the SWCNT metallicity within a SWCNT bundle. We found that a TLL model with the theoretically predicted strength for Coulomb interactions is realized in bundled mixtures of metallic and semiconducting SWCNTs. This demonstrates microscopically that individual metallic SWCNTs within these bundled mixtures can be regarded as isolated metallic SWCNTs. In contrast, although the low-energy spin excitations for bundled metallic SWCNTs still display power-law behavior, which is the hallmark of a TLL state, electronic spin excitations are suppressed by one order of magnitude, with weakened Coulomb interactions, than the bundled mixtures of metallic and semiconducting SWCNTs. This contrasting behavior indicates larger intertube interactions in bundled metallic SWCNTs than in the bundled mixtures. Our findings suggest a mechanism by which the one-dimensional electronic states in SWCNTs can be manipulated.

\begin{acknowledgments}
This work was supported in part by JSPS/MEXT KAKENHI Grant Numbers 15K04601, 16H00919, and 25246006. 
\end{acknowledgments}

%

\end{document}